\documentstyle[prl,aps]{revtex}
\def\ll{\label}
\def\re{\ref}
\def\c{\cite}

\def\r1{(\ref{$1})}

\def\ba{\begin{array}{c}}

\def\ea{\end{array}}

\def\de{\delta}

\def\ov{\over}
\def\ha{{1\over 2}}

\def\l{\left}
\def\l({\left(}
\def\r){\right)}
\def\r{\right}

\def\la{\lambda}
\def\al{\alpha}
\def\be{\begin{equation}}
\def\bc{\begin{center}}
\def\ec{\end{center}}
\def\ee{\end{equation}}
\def\ed{\end{document}}
\def\bea{\begin{eqnarray}}
\def\eea{\end{eqnarray}}

\def\bi{\begin{itemize}}
 \def\ei{\end{itemize}}

\begin{document} 
\title{Refinement  of  Bethe ansatz string  and its alternative  }
\author{
Anjan Kundu \footnote {email: anjan@tnp.saha.ernet.in} \\  
  Saha Institute of Nuclear Physics, \\
 1/AF Bidhan Nagar, Calcutta 700 064, India.
 }
\maketitle
%

\begin{abstract} 

The well known  string solution to  Bethe ansatz  equations is shown to be
inconsistent in its widely accepted form.  A valid refinement demanding
 higher order corrections in subsequent  roots is identified. A new
alternative string solution is proposed for finite long chains, consistent
with the non-Bethe strings observed earlier.

\end{abstract}


Seventy years ego a phenomenal paper  by Hans Bethe \c{bethe31}, opened up a
new direction in physics, namely the theory of
 exactly solvable quantum systems. The pioneering concepts of coordinate
 Bethe ansatz (CBA) and string solution to Bethe ansatz equations (BAE) at
 the thermodynamic
limit, introduced in this paper for the isotropic spin-$\ha$ chain, proved
to be an universal tool for almost all such models discovered in subsequent
years.  Next development, though came rather late,
 solved a number of important problems using the same method. Some of such
milestone works,
 where a wide range of one dimensional models were exactly solved at the
finite chain as well as in the thermodynamic limit may be listed as follows.
 Bose gas interacting through $\de$-function potential \c{LieLin63};
$XXZ$-spin chain \c{YanYan66} along with the application of thermodynamic
Bethe ansatz (TBA)
\c{gaudin71};  repulsive $\de$-function  fermion gas \c{Yan67,Lai71}; 
 1D Hubbard model \c{WuLie}; fully anisotropic  $XYZ$ spin chain \c{baxter}
 etc.
In most of  these models the Bethe string 
  plays a crucial role, especially  in the TBA treatment.
  A systematic
 application of TBA to a variety of models has been considered in details
 in  \c{Schlot}.
Another remarkable achievement 
  is the algebraic formulation of  Bethe ansatz (ABA)
\c{Fadrev,korbook}, which is more suitable for  
 quantum field  models like nonlinear
Schr\"odinger equation  (NLS) \c{Skl1},  sine-Gordon (SG) \c{FaTaSk}
 and Liouville models \c{FadTrik95}.
 Though ABA  differs considerably from its
coordinate
 formulation (CBA), the end equations, e.g.  BAE
 turn out to be equivalent. Therefore the Bethe string 
 solutions may  appear
also in such  
integrable field models forming   quantum  solitons or breathers
 \c{Fadrev,FaTaSk}.
The  widely
 accepted  and frequently quoted form
  of the Bethe string  is  given by  
  \c{Lai71,Schlot,Woy82}
\be 
 \la_l=\la_0 +i (l-{r+1 \ov 2}) +i O
(e^{-\alpha N})
, \qquad  l=1,\ldots,r,
\ll{bs}\ee
with $\alpha$ positive, which is supposed to satisfy the
 Bethe ansatz equations
\be
\left({\la_l+{i \ov 2}\over \la_l-{i \ov 2}}\right)^N=
\prod_{n\neq l}  {\la_l-\la_n+i \over \la_l-\la_n-i}, \qquad  l=1,\ldots,r,
\ll{bel}\ee
 for large  $N$. It is evident from   the above solution
 that exponentially small corrections: $O(e^{-\alpha N}) $ are needed for 
    large but finite values of $N$. Moreover,  the form (\re{bs}) suggests
that these corrections are of the same order for all the roots, i.e. they
should be at least as $\delta_l e^{-\alpha N}.$  Consequently,
  in the  first approximation, i.e. for $N = \infty$, all corrections must
vanish making the string solution exact and
 given  in the obvious form \c{DesLow82,TakhFad81,Izum88}
\be 
 \la_l=\la_0 +i (l-{r+1 \ov 2}), \qquad  l=1,\ldots,r
\ll{bs0}\ee

We however find  surprisingly that 
both the above  well known and well  
 accepted   forms (\re{bs}) and (\re{bs0})  are, strictly speaking,
not 
consistent with the
 BAE (\re{bel}) for higher $r$ ($r\geq 2 $) values. We observe further that
   a fine-tuned form with  carefully defined approximations, where higher and
higher order corrections  appear in the subsequent roots, can only pass
for a valid solution. Such inconsistency in
 the commonly  accepted form of the string solution 
 and  the necessity for its  refinement  perhaps   have
 never been addressed before.
 On the other hand,
 in early eighties a series of papers reported results
 \c{Woy82,DesLow82,BabdeVVia83}, 
 where for long but
finite chains some  non-Bethe string solutions were found to appear.
We intend to   propose here a new string solution
also for finite  long chains, which is consistent with 
the BAE without  finely adjusted
 orders of correction.
 Roots
 within such   strings  may  group together in  
sets of { two}, three and { four} in the thermodynamic limit, providing
 an explanation for the similar
 non-Bethe string  structures 
observed earlier \c{Woy82,DesLow82,BabdeVVia83}.
An important feature of this  alternative string  is that, 
 the spacings between the imaginary parts of its roots, in contrast to   the
 Bethe string, are no longer same for all solutions but depends on the spin
 excitation number $r$. As a consequence the string length does
not grow  linearly with  $r$ as in  the standard case,
but   saturates  as ${r-1 \over r}$ giving only { short strings}.

For concreteness we  consider
 the  Heisenberg
ferromagnet  with nearest neighbor interactions:
  $~~~H_s= -\ha\sum_{n=1}^N \vec \sigma_n \cdot
\vec \sigma_{n+1}
~~. $  We look briefly  into the Bethe ansatz method for solving
 its eigenvalue problem   \c{bethe31} to see 
  how BAE (\re{bel}) arise, 
 before concentrating on the string solution (\re{bs}).
  The  Bethe ansatz  for the eigenfunction in 
 $r$-down spin state:
   $\Psi_r=\sum_{\{m_i\}} a(m_1,m_2, \ldots,m_r)
|m_1,m_2,\ldots,m_r>$  is chosen as
\be
 a(m_1,m_2,\ldots,m_r)=\sum_P \exp{i[\sum_{l=1}^r k_{P_l} m_l+\ha \sum_{l<n}
\phi_{P_l,P_n}]},
\ll{ba}\ee
where $P$ is any permutation of the numbers $1,2,\ldots,r$ and $P_l$ is 
the number that replaces $l$ under this permutation.
It is  easily seen that if none of these spin excitations  are adjacent, they
become noninteracting and the Schr\"odinger equation $H \Psi_r=E_r\Psi_r$
leads to 
\be
 E_ra(m_1,m_2, \ldots,m_r)=
\sum_{j=1}^r [a(\ldots,m_j+1, \ldots)+
a(\ldots,m_j-1, \ldots) -2 a(\ldots,m_j, \ldots).
\ll{eve1}\ee
The ansatz (\ref{ba}) representing 
a superposition of plane waves is clearly  an
  exact solution of (\ref{eve1}) 
 for arbitrary phases.
 However as soon as any two of the spins become adjacent,
say $  m_{l+1}=m_l+1,$   interaction sets in making
the  equation  different from (\re{eve1}). These two sets of equations 
can only become consistent if the    condition on the amplitude
$~~ 2a( \ldots,m_l,m_{l+1},\dots)=
a(\ldots,m_l,m_{l}, \ldots)+
a(\ldots,m_l+1,m_{l}+1, \ldots).~~$
is satisfied.
This in turn fixes  the phase  through pseudomomenta as
$~~cot \phi_{ln}=\ha(cot {k_l \over 2}-cot {k_n \over 2}), \ \ \ -\pi \leq 
\phi_{ln} \leq+\pi~~ $ with the 
obvious antisymmetry  property $\phi_{nl}=-\phi_{ln}. $
The next problem is to identify the equations for determining 
 the  variables $k_l, \ 
l=1,2,\ldots, r$, for which 
we use the periodic boundary condition $ ~~
a(m_1,m_2, \ldots,m_r)=a(m_2, \ldots,m_r, m_1+N) ~~  $ and involve 
 ansatz (\ref{ba}) to get  an important set 
of equations 
\be
e^{ik_l N}=e^{i\sum_{n\neq l} \phi_{ln}}, \ \  l=1,\ldots, r, 
\ll{be}\ee
known as BAE.
 Introducing    
rapidity variables as
 \be \la _l= \ha cot {k_l \over 2}, \ \ \ \  \mbox{giving} \  \
e^{i \phi_{ln}}={\la_l-\la_n+i \over \la_l-\la_n-i} \ll{map}\ee 
 one can easily rewrite BAE (\re{be})  in the form (\re{bel}).
It is worth noting that in the  
ABA method the BAE appear automatically in the form
 (\re{bel}),
where 
 $\la_l$'s correspond to fixed values of the  spectral parameter. 
The solutions for $\la_l$
to   BAE (\re{bel}) must be  equivalent to the solutions $k_l$
for 
 (\re{be}), though their ranges  are different.
The total number of  solutions of BAE, as shown in  \c{bethe31}, 
must be given by real as well as complex roots. 
  To find the complex solutions for $k_l=u_l+i v_l$, we  notice that
for positive (negative) values of $v_l$ the LHS of
 (\re{be}) exponentially vanishes (diverges) 
for large $N$. Therefore at least one of the $ \phi_{ln}$'s must have 
positive (negative) imaginary part
dominatingly large  as $O(N)$. 
The well known 
$r$-Bethe string solution, which is   
supposed to fulfill this criterion is given  in the form
 (\re{bs}) or (\re{bs0})
for variables $\la_l,$ as mentioned above. 

Let us however check the validity of this solution for all values of $r$
  by direct insertion into BAE (\re{bel}).
In the 
simplest case of $r=2$ we get from (\re{bs0}) in the first
approximation:
$ \la_1=\la_0 -i {1 \ov 2},
 \la_2=\la_1^* ,$ forming a complex conjugate pair.
 Therefore it is enough to check Eqn (\re{bel}) only for  $l=1$, where
$ \mbox {LHS}= \left({\la_0\ov \la_0^2 -{i }}\right)^N$ gives 
a vanishing term
 at  $N = \infty$. This   is easily seen by taking its modulus. 
 $ \mbox {RHS}= {\la_1-\la_2+i \over \la_1-\la_2-i}$ 
on the other hand  vanishes due to $\la_1-\la_2={-i},$
   proving thus the $2$-string as a valid  solution.   
  For  $r=3$
 with $\la_2=\la_0$ real and
 $\la_1= \la_3^*=\la_0 -i,$
we have to check equations for $l=1,2$.
 Here again  for the  neighboring pairs 
   $\la_1-\la_2=\la_2-\la_3= -i$ and consequently,
 the equation for $l=1$ holds
 on similar grounds as  above at large $N$.
 For $l=2$ on the other hand, 
  zeros appearing in the RHS from 
the terms with $n=1$ and $n=3$ get  canceled
  giving a 
 nonvanishing   term.  This however does not  
 lead to any   contradiction,
since  $\la_2$ being   real
 the LHS is a pure phase and does not vanish 
 for any $N$.

Since the Bethe string  in the form (\re{bs}) 
is  valid, as shown above,  
 for  $r=2$ and  $r=3,$ it is probably
natural to assume that  it will be valid  for arbitrary
$r,$ { by analogy}.
 This assumption is generally  made in most of the   works 
  {in proving}  Bethe string
solution for higher $r$ \c{TakhFad81,Izum88}.    
We   show however that this assumption 
 is not true in general and for 
demonstration consider the case  $r=4$ in some detail, where
the Bethe string obtained from (\re{bs}) is of the form
\be
 \la_{1}=\la_0 - i {3 \ov 2} +i O(e^{-\al N}), \ \ 
  \la_{2}=\la_0 - i {1 \ov 2} +i O(e^{-\al N}), \ \ \
 \mbox {with} \ \  \la_{4}=\la_1^*, \  \la_{3}=\la_2^*, \ll{4s}\ee
and one has to check only for 
the independent solutions with $l=1,2.$  
Defining for convenience a function 
$V_s(x)=  {x+i s \over x-is}$ with the obvious property
\be \ V_1\left(\mp i +O(e^{-\al N})\right)~=~O(e^{ \mp \al N}), \ll{prop}\ee  
we see  for $l=1$ that  the LHS of  (\re{bel}): $~V_{\ha}^N(\la_1)= 
\left( {\la_0-i 
 \over \la_0-2i}\right)^N$ vanishes exponentially for large $N,$
 while the RHS is a product of
 three
factors $  ~ V_{1}(\la_1-\la_2)V_{1}(\la_1-\la_3)V_{1}(\la_1-\la_4).$ 
Since only  neighboring rapidities can give $\la_1-\la_2=-i+O(e^{-\al N}),$
we conclude immediately that the RHS also  vanishes exponentially
due to the single factor $V_{1}(\la_1-\la_2)=
O(e^{-\al N})$ and the remaining  giving  finite contributions.
This clears  the string solution for $l=1$.
However, as we see below, the contradiction starts from $l=2,$.
  The LHS in this case:   $V_{\ha}^N(\la_2)= 
\left( {\la_0
 \over \la_0-i}\right)^N$   is again exponentially small and 
the RHS contains 
 three
factors $~   V_{1}(\la_2-\la_1)V_{1}(\la_2-\la_3)V_{1}(\la_2-\la_4).$ 
However the important point   is that, there are now two 
 neighboring roots of $\la_2$   contributing as $ 
\la_2-\la_3=-i+O(e^{-\al N})$ and $
\la_2-\la_1=i+O(e^{-\al N})$ and consequently  using  property (\re{prop})
 we get $V_{1}(\la_2-\la_3)=
O(e^{-\al N}),$ while $V_{1}(\la_2-\la_1)=
O(e^{\al N})$ and $ V_{1}(\la_2-\la_4)$ as a finite term. Therefore
the important conclusion we arrive at is that, all singularities
 in the terms
of  RHS
 get mutually 
canceled leaving only a finite contribution even at $N \to \infty$,
contrasting the LHS.
It is apparent now  from the above reasoning 
that this inconsistency  would  
 always appear for higher $r,$ in the  Bethe string solution  in its
accepted form (\re{bs})  for  all its roots $\la_l, $ 
with $l=2,\ldots, r-1,$  having 
 two nearest  neighbors, giving 
\bea
\mbox{LHS}&=&V_{\ha}^N(\la_l) \sim O(e^{-v_l N}), \ll{L} \\
 \mbox{while} \qquad ~~~~~ ~~~~~~~~~~~~
\qquad \qquad   \qquad  \qquad    \mbox{RHS}
&\sim &    V_{1}(\la_{l}-\la_{l+1}) V_{1}(\la_{l}-\la_{l-1})~ =
~O(e^{-\al N})
O(e^{\al N}) \sim O(1).
\ll{LnR}\eea
 Only for the   end roots  having single
neighbors or for  roots with real values, the form (\re{bs})
 remains consistent.
 Since
for $r=2,3$ all roots fall into this category, they give valid string
solutions.
 
The next natural question is  how to refine
 the structure of  (\re{bs}), so that we can get valid Bethe string 
 for all values of spin excitations $r.$  For answering
 this question we see from (\re{LnR}) that  the 
reason for  failure of the string form (\re{bs}) is that, it allows 
  the same
order of correction $O(e^{-\al N})$   for all its 
roots $\la_{l}$. Therefore, the 
refined   form for  the Bethe string, consistent  for arbitrary $r$
 should be given as
\be 
 \la_l=\la_0 +i (l-{r+1 \ov 2}) +i O(e^{-\alpha_l N})
,\  \ \mbox {and} \  \ \la_{r+1-l}=\la_l^*, ~~~  l=1,\ldots, s,  
\ll{bsr}\ee
with $r=2s$  for even and $r=2s+1$ for odd $r$ with 
real $~~\la_{s+1}=\la_0 $.
 The essential
point for the validity of this refined string   is that, 
the exponential orders in 
 (\re{bsr}) must be fine-tuned as a strictly growing sequence 
\be
0<\alpha_1 <  \cdots <\alpha_l<\alpha_{l+1}<\cdots<\alpha_s,
\ll{grow}\ee
and more precisely they should be given by the recurrence relation
\be
\alpha_l-\alpha_{l-1}= v_l ,~~~~ \ \ \mbox{where}~~ ~~~ v_l=\ha ~ln
(1+\kappa_l) >0 ,
\ll{rel}\ee
with $ \kappa_l={r+1-2l \ov
\la_0^2+({r \ov 2}-l)^2} > 0 $ for $ l=1,\ldots, s.$
To check the validity of the refined  form (\re{bsr}) we insert it 
again  in
 BEA (\re{bel}) and notice that  the LHS $\sim O(e^{-v_l N})$
  at large $N$ similar to (\re{L}), while the RHS is now given by
\bea
 \mbox {RHS}
&\sim &    V_{1}(\la_{l}-\la_{l+1}) V_{1}(\la_{l}-\la_{l-1}) 
 =\left( O(e^{-\al_l N})-O(e^{-\al_{l+1} N})\right)
\left(O(e^{-\al_l N})-O(e^{-\al_{l-1} N})\right)^{-1} \nonumber \\
&\approx & O(e^{-\al_l N})O(e^{\al_{l-1} N}) \sim O(e^{-v_{l} N}),
\ll{R}\eea
     which proves the claim. 
 In deriving the expressions in 
(\re{R}) we have  used the relation (\re{rel})
and approximated  $O(e^{-\al_l N})-O(e^{-\al_{l+1} N}\approx 
O(e^{-\al_l N}) $ and $O(e^{-\al_l N})-O(e^{-\al_{l-1} N}\approx 
O(e^{-\al_{l-1} N})$ neglecting higher order smaller terms due to
  strict inequalities (\re{grow}). It is important to note that
since  all higher orders of correction are present simultaneously  in  
(\re{bsr}), it is not principally possible  to approximate 
this solution order by order for   $r \geq 4$ cases.
 For example, one may  naively  consider 
  the first approximation by putting all corrections  $O(e^{-\al_l N})=0 $
 in (\re{bsr}) reducing it to the form (\re{bs0}), which is however
not a valid solution for  $r \geq 4,$  as we have seen above. 
In any approximation,  one can 
at best neglect   the highest order of
correction by putting  $O(e^{-\al_{s N}})=0 $, keeping however all other
corrections as in (\re{bsr}). The limit $N=\infty$
should be inserted in the solution 
only at the end after performing  all calculations,  even in the first
approximation. For $r=2, 3$  however, since  $s=1,$  
 we can neglect all corrections in (\re{bsr})     
in the first approximation  and get (\re{bs0}) as the consistent solution.
In fact for these cases only the form (\re{bsr}) coincides with  
(\re{bs}).

It is fair to  mention here that, though the 
Bethe string  in the  form 
 (\re{bsr}) refined through 
 (\re{grow}), (\re{rel}), as far as we know,  has not appeared
  in any of the earlier
works,
the form presented by Gaudin \c{gaudin71} agrees qualitatively 
with (\re{bsr}), though
the requirement (\re{grow}), (\re{rel}) necessary for the validity
of the solution was not specified in his paper.
Note that the complex  string  solution for 
pseudomomentum $k_l$ equivalent to  (\re{bsr}) should be given 
through the mapping (\re{map}), from which   it is  clear that
 in the  corresponding phase  
$i\sum_{n \neq l} \phi_{ln}$ appearing  in (\re{be}),  the terms
 $ \phi_{ll\pm 1}$ must have large imaginary parts  having 
corrections like $i\beta_l N$ and $-i\beta_{l-1} N,$
 with the   inequalities 
\be
0<\beta_1 <  \cdots <\beta_{l-1}<\beta_{l}<\cdots <\beta_s, ~~~
\beta_{l}-\beta_{l-1}=v_l,
\ll{growp}\ee
  reflecting (\re{grow}), (\re{rel}).
  Bethe  \c{bethe31}
suggested also that {\it at least one of the $ \phi_{ln}$  has to have a
very large imaginary part  of   $O(N)$}.
 However the essential requirement (\re{growp})
 for the consistent higher $r$ strings was not  emphasized and
 (\re{bs0}) was taken as the solution for arbitrary $r$ in \c{bethe31}
(see Eqn (30)).
 We have shown however
  that one can not get any solution in
the form (\re{bs0})
 for $r \geq 4$, since the
 corrections   can not be
 neglected from the beginning, 
 even in the first approximation.

Thus we have identified a consistent
 but complicated form  (\re{bsr})  for the Bethe string,
which demands  simultaneous
involvement of all higher order corrections (up to $s$) 
for its
validity.        However such fine-tuned structures growing linearly with
$r$   might be on  physical grounds less  
 probable to appear    
      in low lying excitations.
Earlier  findings of  short strings with 
non-Bethe  nature seem  to support such arguments.
 This   therefore motivates us    to find  some new type of string solution 
with desirable properties 
like shorter string length; possibility of order by
order  approximation and finally 
possible explanation of      
  non-Bethe  strings observed  earlier.
 The  alternative string  we propose has the form
 \be 
 \la_l=\la_0 +i {1\ov s} (l-{r+1 \ov 2}) +i O
(e^{-\alpha N}),  \ ~~ ~  l=1,\ldots, r.  
\ll{as}\ee
 Note that
   the
corrections appearing in  all its roots have   the same  order
and the
 spacing between the neighboring roots 
 $d(1)={ i\ov s}$ varies inversely  with $r$.  
 In contrast to the Bethe string  with constant spacing
$d(1)=i,$  
  the  
 string (\re{as})  with more roots is more
dense along the imaginary axis and its    
 length: $L(r)= {2({r-1 }) \ov r},$
measuring the distance between  end roots, is bounded between
 $ 1\geq L(r) \geq 2.$  Recall that for 
   Bethe 
 string   $L(r)= {r-1}.$ 

For checking  the validity of  (\re{as})
we put it directly  in BAE (\re{bel}) and   observe that 
while the LHS is 
similar to (\re{L}) giving exponentially vanishing term    
$V_{\ha}^N(\la_l) $ for large $N$, the RHS behaves differently 
 from the standard case (\re{LnR}). Since now  $d({s})=i$, 
the pairing partner
 having  the crucial difference $\la_l- \bar \la_l=-i +
O(e^{-\al N})$ 
is given by
 $\bar \la_l= \la_{l+s}$. It is also  seen easily that
  in  this case for every complex root $\la_l$
 such pairing can occur only once, i.e. 
 there can be no 
other root except $\bar \la_l$ giving either $i$ or $-i$.
 This is in sharp contrast with the standard Bethe string  (\re{bs}),
where singularities can come  from  two  neighboring roots.
Therefore unlike  (\re{LnR}),  we get for (\re{as})
 $ \mbox {RHS}~
\sim   ~  V_{1}(\la_{l}-\la_{l+s})  =
O(e^{-\al N}),$ i.e. an exponentially vanishing term like LHS.
It is also worth noting that since the alternative string (\re{as}) allows  
 order by order approximation one can consider  the first
approximation by putting $N=\infty,$ when the correction term $O(e^{-\al N})$
 drops out from all roots of  (\re{as}),
 making it an exact solution.  
Recall again that this is not possible for a  consistent 
 Bethe string given in the  form (\re{bsr}).

From the above arguments it is clear that 
for each
   $ \la_l$  in  
(\re{as})  its partner  $\bar \la_l=\la_{l+s}$ and its conjugate
  $ \la_l^*=\la_{r+1-l}$
are  also included in the solution and since the partner of its partner and
conjugate of its conjugate is the root itself, the group of four 
$\{\la_l, \bar \la_l, \la_l^*, \bar \la_l^*\}$  forms a close unit, 
 being  the decisive   contributors to the   equation at
 large $N$ and thus  becomes almost independent entries.
Therefore all the $r$ roots can be thought of to be  broken up
into units of ${\it four}$ at the thermodynamic limit,
 providing an  explanation of the 
quartet formations found earlier.
However when the number 
 $r$ is not divisible by $4$, doublet and triplet groups can also occur.
For $l={r+2 \ov 4}$ when
$r$ even and for  $l={r+3 \ov 4}$ when  $r$ odd, one gets $\bar \la_l=
 \la_l^*, $ and consequently it can group only
 as a unit of {\it two}, forming a two string.
 For odd $r$,  $\la_{s+1}=\bar \la_1=\bar \la_1^*$ is real
and gives a triplet  $\{\la_{1},\bar \la_1, \la_1^*\}$,
forming a $3$-string.
So the general rule is that,  in the thermodynamic limit  $r=4  n$ results
 $n$ quartets, 
$r=4n+1$  gives $n-1$ quartets, a  doublet and a triplet, 
$r=4n+2$, yields   $n$ quartets and a  doublet and  the case 
$r=4n+3$  groups into
 $n$ quartets and a  triplet.
Thus within the framework of a new kind of short-lengthed $r$-string
 (\re{as}), we can 
provide a possible explanation of the non-Bethe string structures that have
been observed for long but finite chains 
\c{Woy82,DesLow82,BabdeVVia83}. For
 the Bethe string  on the other hand, each  root generally 
has two
partners as its nearest neighbors, each of which in turn has its 
own two   partners. Therefore the Bethe string 
can not close into any shorter
 substrings. 

 For the    alternative $r$-string 
   (\re{as})
the  total momentum   and the excitation  energy
 may be given by
$~~P^{(as)}_r=\sum_{j=1}^s p_j,~~~ \mbox {with}~~~~  
cot ~{p_j\ov 2}= {\la_0 \ov g_j},~~~$ and
\be
E^{(as)}_r={\partial P \ov \partial \la_0}=
\sum_{j=1}^s {2 g_j \ov \la_0^2+g_j^2},
\ll{E}\ee
where   the factor  $g_j=g_1-{1 \ov s}(j-1)$, with $g_1={3 \ov 2}$
 for odd  and
 $g_1={3 \ov 2}-{1\ov 2s}$ for even values of $r.$   

We  note here that though
 the eigenfunction
 (\re{ba}) for   solution (\re{as})  does not  vanish
exponentially  in all directions and therefore does not give fully bound
states, it should  nevertheless be normalizable, since this solution  
can be considered only  for long
but finite chains with periodic boundary conditions (see \c{Woy82}).
We believe that, the energy (\re{E})
 of  our  string  is always  
lower than the energy of $r$-free magnons with rapidity $\la_0$: $rE_1=
 ={r \ov \la_0^2+{1 \ov 4}}.$ Though we could not prove this conjecture in
general, we  checked it for a number of cases. Few of such sample cases are
 $r=4,5,7,9,10$ with  $\la_0=1.0,$  which    gives $E^{(as)}_r
=1.94, 1.92, 2.9, 3.9, 4.8$
and $r   E_1=3.2, 4.0, 5.6, 7.2, 8.0$, correspondingly.
Any  new string    solutions like  (\re{as}) 
or  non-Bethe  excitations  observed earlier,
 if they exist,  should 
   possibly arise  from the interactions of different Bethe strings at the
thermodynamic limit,
 since the later form a complete set.

Precise formulation of the   valid  Bethe string (\re{bsr}) with 
(\re{grow}), (\re{rel}) and presentation of a new string
(\re{as}) are the main results of this paper.
 Solution (\re{as}) can  be generalized to 
strings with nonuniform spacing, since at the thermodynamic limit it
 splits functionally into smaller groups with not
 much relevance on their relative distances.
Non-Bethe macroscopic string 
 with nonuniform spacing has been considered recently \c{shastry00}. 
The results obtained here
 for the simplest case of $XXX$-spin chain can be
extended also to  other integrable 
 models  allowing Bethe string solutions,
 like $XXZ$-spin chain, 1D Hubbard model,
supersymmetric $t-J$ model etc. and field models 
 like NLS, SG, Liouville model.
Therefore, application of the  TBA  by  using the 
alternative string proposed here to the models 
investigated earlier with the use of 
Bethe strings \c{Schlot}  would be an interesting problem.

Acknowledgment: I would like to thank  A. Chatterjee, B. Basu-Mallick 
,  P. Mitra and G. Bhattacharya for fruitful discussions, 

 \end{document}